\documentclass[12pt,a4paper]{article}
\newcommand\be{\begin{equation}}
\newcommand\ee{\end{equation}}
\newcommand\bea{\begin{eqnarray}}
\newcommand\eea{\end{eqnarray}}

\newcommand{\fatalpha}{{\bf \alpha \kern -0.44em \alpha}}
\newcommand{\fatsigma}{{\bf \sigma \kern -0.54em \sigma}}
\newcommand{\tpchi}{{\bf \chi \kern -0.35em \chi}}
\newcommand{\llambda}{{\bf \lambda \kern -0.45em \lambda}}

\begin{document}
\title{Quantum Search for Zeros of Polynomials}
\author{Stefan Weigert \\
HuMP - Hull Mathematical Physics\\ Department of Mathematics,
University of Hull, UK-Hull\\
\tt s.weigert@hull.ac.uk}
\date{May 2003}
\maketitle
\begin{abstract}
A quantum mechanical search procedure to determine the real zeros
of a polynomial is introduced. It is based on the construction of
a spin observable whose eigenvalues coincide with the zeros of the
polynomial. Subsequent quantum mechanical measurements of the
observable output directly the numerical values of the zeros.
Performing the measurements is the only computational resource
involved.

\end{abstract}

%
%
\subsection*{Introduction}
Quantum mechanical measurements are a computational resource.
Various quantum algorithms use projective measurements at some
stage or other to determine the period of a function e.g.
\cite{shor94}. In \cite{nielsen01}, projective measurements have
been assigned a crucial role for a particular scheme of universal
quantum computation which requires measurements on up to four
qubits. Related schemes have been formulated based on measuring on
triples and pairs \cite{fenner+01}, and finally on pairs of qubits
only \cite{leung02}. Measurements are also an essential part of
Grover's search algorithm in order to actually read the result of
the computation \cite{grover97,farhi+98}.

A conceptually different strategy has been applied to propose a
special-purpose machine which is capable to diagonalize any
finite-dimensional hermitean matrix by genuine quantum means, i.e.
quantum measurements \cite{weigert01,weigert03}. In this approach
of {\em quantum diagonalization}, a hermitean matrix is considered
as a quantum mechanical observable of an appropriate one-spin
system. Projective measurements with a generalized Stern-Gerlach
apparatus provide directly the unknown eigenvalues of the matrix
which solves the hard part of the diagonalization. In this paper,
it will be shown how to find the real zeros of a prescribed
polynomial in a similar way, using quantum mechanical
measurements.
\subsection*{The quantum search procedure}
Consider a polynomial of degree $N$ which is assumed to have $N$
real zeros $\zeta_n$,
\be
P(x) = \sum_{n=0}^N p_n x^n \, , \quad p_n \in {\sf I R}\, ,
\quad p_N = 1 \, .
\label{Npoly} \ee
The assumption $p_N = 1$ is not a restriction since two
polynomials $Q(x)$ and $Q(x)/c , c \neq 0$, have the same zeros.
The quantum procedure to identify the zeros $\zeta_n$ of $P(x)$
consists of two steps. First, one needs to find a hermitean
companion matrix ${\sf C}$ of the polynomial $P(x)$. By
construction, its eigenvalues coincide with the unknown zeros of
the polynomial $P(x)$. Second, one determines the eigenvalues of
the matrix {\sf C} by a method inspired by the quantum
diagonalization of a hermitean matrix. Effectively, they are
obtained by measuring the eigenvalues of a quantum mechanical spin
observable ${\hat C}$ with matrix representation {\sf C}.

\subsection*{Hermitean companion matrix of a polynomial}
It is straightforward to calculate the characteristic polynomial
$P_{\sf M}(\lambda)$ of an $(N \times N)$ matrix {\sf M}:
\be
\det \left( {\sf M} - \lambda {\sf E} \right)
    = P_{\sf M}(\lambda) \, ,
\label{charMpoly} \ee
where {\sf E} is the $(N \times N)$ unit matrix. The polynomial
$P_{\sf M}(\lambda)$ has degree $N$, and its zeros coincide with
the eigenvalues $\mu_n, n=1, \ldots, N$, of the matrix {\sf M}.
Hermitean matrices have real eigenvalues only, hence all zeros of
its characteristic polynomial are located on the real axis. The
inverse problem reads:
\begin{itemize}
    \item Given a polynomial $P(x)$ of degree $N$  with real
    zeros, find a {\em hermitean} matrix {\sf C} such that its
     characteristic polynomial is $P(x)$.
\end{itemize}
The matrix {\sf C} is known as {\em companion} matrix of the
polynomial $P(x)$. Obviously, the companion matrix should be
determined {\em without} reference to the roots of the given
polynomial. If {\sf C} is one solution of the inverse problem,
then the matrices ${\sf C}_{\sf U} = {\sf U} {\sf C} {\sf
U}^\dagger$ provide solutions as well, where {\sf U} is any
unitary $(N \times N)$ matrix. The difficult part of the inverse
problem lies in the requirement to find a {\em hermitean} matrix
{\sf C}: it is easy to specify a non-hermitean companion matrix
${\sf C}_0$ of the polynomial (\ref{Npoly}), {\em viz.},
\be
{\sf C}_0 = \left(
\begin{array}{cccc}
      0 &    1 &        &          \\
 \vdots &      & \ddots &          \\
        &      &        &       1  \\
   -p_0 & -p_1 & \cdots & -p_{N-1} \\

\end{array}
\right) \, .
\label{nonhcompanion} \ee
Partial solutions of the problem to find hermitean companion
matrices have been obtained in \cite{fiedler90}. A complete and
constructive solution can be found in \cite{schmeisser93}, where a
tridiagonal companion matrix {\sf C} is  specified in terms of the
coefficients $p_n, n=0,1, \ldots,N$. Explicitly, a polynomial
$P(x)$ of degree $N$ with $N$ real zeros can be written in the
form
\be
P(x) = (-)^N \det \left( {\sf C} - x {\sf E}\right) \, ,
\label{polyasdet} \ee
where the matrix {\sf C} is tridiagonal and real symmetric, hence
hermitean:
\be
{\sf C} = \left(
\begin{array}{ccccccc}
-q_1(0)   &\sqrt{d_1}&          & & & &                   \\
\sqrt{d_1}&-q_2(0)   &\sqrt{d_2}& & & &                   \\
            &\sqrt{d_2}& -q_3 (0)   &\ddots & & &         \\
            &          & \ddots   &\ddots & & &           \\
         & & & &   -q_{n-2}(0)&\sqrt{d_{n-2}}&              \\
         & & & &\sqrt{d_{n-2}}&-q_{n-1}(0)   &\sqrt{d_{n-1}}\\
         & & & &              &\sqrt{d_{n-1}}&      -q_n(0)
\end{array}
\right) \, .
\label{companion} \ee
The nonnegative numbers $d_k, k = 1, 2, \ldots, N-1$, and the
polynomials $q_n (x), n = 1,2, \ldots, N$, are generated when
applying a {\em Modified Euclidean Algorithm} \cite{schmeisser93}
to the polynomial $P(x)$. The operations required to determine the
matrix elements of ${\sf C}$ are ($\imath$) repeated division of
polynomials, ($\imath \imath$) evaluation of specific polynomials
at $x=0$; ($\imath \imath \imath$) taking square roots of numbers
$d_k$. As a corollary, the Modified Euclidean Algorithm checks
whether the given polynomial has real zeros only: if any of the
numbers $d_k$ is found to be negative, $P(x)$ can not have real
zeros only. For completeness, the algorithm is sketched in the
Appendix.

\subsection*{Quantum search for eigenvalues of hermitean matrices}
Four steps are necessary to find the eigenvalues of a given
hermitean matrix ${\sf C}$ with $N$ different eigenvalues by means
of quantum measurements. Here an outline of this approach will be
given only; for details about the procedure for $(N \times N$)
matrices readers should consult \cite{weigert01}, while it is
illustrated for $(2 \times 2)$ matrices in \cite{weigert03}.

The matrix must be ($\imath$) written in {\em standard form};
next, it is interpreted as ($\imath \imath$) matrix representation
of a unique {\em observable} ${\hat C}$ of a quantum spin; this
observable can be measured by ($\imath\imath \imath$) a specific
{\em apparatus} which needs to be identified and built; finally,
the apparatus is used to generate the eigenvalues by ($\imath v$)
actually {\em measuring} the observable ${\sf C}$.

\begin{enumerate}
  \item[($\imath$)] {\em Standard form of ${\sf C}$:}
      Write the hermitean $(N\times N$) matrix ${\sf C}$ as
      a combination of linearly
      independent hermitean {\em multipole} operators
      ${\sf T}_\nu, \nu= 0, \dots ,N^2-1,$
\be
{\sf C} = \sum_{\nu=0}^{N^2-1} {c}_{\nu} {\sf T}_{\nu} \, , \qquad
{c}_{\nu} = \frac{1}{N}\mbox{ Tr } \left[{\sf C} \, {\sf T}_{\nu}
\right] \in {\sf I R} \, .
\label{expandgen} \ee
There are $N^2$ self-adjoint multipole operators ${\hat T}_\nu =
{\hat T}^\dagger_\nu$ with matrix representations ${\sf T}_{\nu}$,
and they form a basis in the space of hermitean operators acting
on an $N$-dimensional Hilbert space ${\cal H}$ \cite{swift+80}.
These operators consist of all traceless symmetric products of up
to $N$ spin components ${\hat {\bf S}}= ({\hat S}_x , {\hat S}_y ,
{\hat S}_y) $, plus the identity operator. Two multipoles are
orthogonal with respect to a scalar product defined as the trace
of their product: $(1/N) \mbox{ Tr } \left[{\hat T}_{\nu} {\hat
T}_{\nu'} \right] = \delta_{\nu\nu'}.$

\item[($\imath\imath$)] {\em Identification of an observable:}
        On the basis of the expansion (\ref{expandgen}) interpret
        the matrix ${\sf C}$ as representing an observable
        ${\hat C}$ for a spin with quantum number $s= (N-1)/2$:
\begin{equation}\label{observableha}
{\hat C} = C ({\hat {\bf S}})
         = \sum_{\nu=0}^{N^2-1} c_{\nu} {\hat T}_{\nu} \, ,
\end{equation}
thinking of the multipoles as functions of the spin components,
${\hat T}_\nu = T_{\nu} ({\hat {\bf S}})$.
\item[($\imath\imath\imath$)] {\em Setting up a measuring device
  for $\hat C$:}
   Swift and Wright have shown in \cite{swift+80} how to
  devise a {\em generalized Stern-Gerlach apparatus} which measures
any spin observable ${\hat C}$. The construction generalizes the
traditional Stern-Gerlach apparatus which measures the spin
component ${\bf e}_n \cdot {\hat {\bf S}}$ along a direction
specified by a unit vector ${\bf e}_n$. Setting up this device
requires to create arbitrary static electric and magnetic fields
in the laboratory, consistent with Maxwell's equations. The
procedure is made explicit in \cite{swift+80}. Once constructed,
the apparatus will split an incoming beam of particles with spin
$s$ into $N=(2s+1)$ subbeams corresponding to the eigenvalues of
$\hat C$. The working principle is equivalent to that of a
standard Stern-Gerlach apparatus where $\hat C \equiv {\hat S}_z$.
\item[($\imath v$)] {\em Determination of the eigenvalues:}
Prepare a spin $s$ in a homogeneous mixture ${\hat \rho}_0 = {\hat
I}/(2s+1)$. When carrying out measurements with the apparatus
associated with $\hat C$, the output of each individual
measurement will be one of the eigenvalues $\zeta_n$ of the matrix
${\sf C}$. The actual values of the eigenvalues can be determined
from the amount by which the particles in each subbeam are
deflected from the straight line of flight. After sufficiently
many repetitions, all eigenvalues $\zeta_n$ will be known. Since
each eigenvalue occurs with probability $1/N$, the probability
{\em not} to obtain one of the $N$ values decreases exponentially
with the number of runs.
\end{enumerate}
By construction, the numbers $\zeta_n$ obtained in the last step
coincide with the zeros of the polynomial $P(x)$, and one can
write
\be
P(x) = \prod_{n=1}^N \left( x - \zeta_n \right) \, .
\label{product} \ee
The zeros $\zeta_n$ have been obtained by a genuinely quantum
mechanical method.

\subsection*{Conclusions}
It has been shown that one can build a quantum mechanical
special-purpose machine which is capable to output the roots of a
polynomial of degree $N$. The underlying working principle is to
perform appropriate quantum mechanical measurements using a
generalized Stern-Gerlach apparatus.

The possibility to extract information from individual quantum
mechanical measurements may have implications for the
interpretation of quantum mechanical states. Quantum root
extraction of polynomials seems to strengthen the {\em individual}
interpretation \cite{jammer74}: a single run of a measurement (with only one
individual quantum system involved) provides information about one
(randomly selected) zero of the polynomial; hence, no ensemble of
identically prepared systems is required to obtain a useful answer
from the experiment.

Instead of repeating the experiment $M \gg N$ times, one can
imagine to run $M$ identical experiments simultaneously, each
generalized Stern-Gerlach apparatus being tuned to search for the
zeros of the same polynomial $P(x)$. If the input state is an
$M$-fold direct product of the homogeneous mixture ${\hat
\rho}_0$,
\be
{\hat \rho}_0 \otimes \ldots \otimes {\hat \rho}_0 \, ,
\label{mfold} \ee
the resulting `parallel' quantum search would, with large
probability, produce {\em all} zeros $\zeta_n$ at one go. In a
sense, quantum mechanics is able to point almost instantaneously
at the zeros of a given polynomial without any software program
running.

\subsection*{Appendix: Modified Euclidean Algorithm}
Given a polynomial
\be
P(x) =  x^N + p_{N-1} x^{N-1} + \ldots + p_0 \, , \quad p_n \in
{\sf I R}\, ,
\label{Npolyalg} \ee
the Modified Euclidean Algorithm \cite{schmeisser93} defines
recursively a number of polynomials $P_k(x)$, $k=2,3, \ldots,
N-1$, of smaller degrees, and it generates other polynomials
$q_k(x)$ and numbers $d_k$ which are required to define the
hermitean companion matrix ${\sf C}$. Start with
\be
P_1(x) = P(x) \, \qquad P_2 (x) = \frac{1}{N} \frac{dP(x)}{dx} \, ,
\label{startalg} \ee
and iterate the following steps. Divide $P_{k}(x)$ by $P_{k+1}(x)$,
\be
P_k(x) = q_k (x) P_{k+1} (x) - r_k (x) \, ,
\label{division} \ee
with a remainder $r_k (x)$ which may either be different from or
equal to zero. Denote by $d(Q(x))$ the coefficient of the highest
power of the polynomial $Q(x)$: for example, $d ( P(x) ) = 1$.
Define a polynomial $P_{k+2} (x)$ and the number $d_k$ according
to

\begin{itemize}
    \item[($\imath$)] $r_k(x)\neq 0:  P_{k+2} (x) =
    \frac{r_k(x)_k}{d_k} \mbox{ and } d_k = d(r_k(x)) \, ;$
      \item[($\imath\imath$)] $r_k(x) = 0:  P_{k+2} (x) =
                       \frac{dP_{k+1}(x)/dx}{d(dP_{k+1}(x)/dx)}
\mbox{ and } d_k = 0 \, .$
\end{itemize}

The algorithm terminates if $P_{k+1} =1$, defining $q_{k}(x)=
P_k(x)$. Otherwise the procedure is repeated with $P_{k+1} (x)$
and $P_{k+2} (x)$ in (\ref{startalg}) and it will generate
$q_{k+1}$ and $d_{k+1}$, etc.

\end{document}